\documentclass[pra,
superscriptaddress,
nofootinbib,%
raggedfooter,%
nopreprintnumbers,showpacs,thigthen,twocolumn,floatfix,amssymb,amsbsy,amsmath]{revtex4-1}%
\usepackage[pdftex]{graphicx}
\usepackage{palatino}
\usepackage[utf8]{inputenc}
\usepackage{enumitem}
\def\physrep{Phys. Rep.\ }
\def\pra{Phys. Rev. A\ }

\def\prd{Phys. Rev. D\ }

\def\nphysa{Nucl. Phys. A\ }

\def\vec#1{\boldsymbol{#1}}
\usepackage{color}
\begin{document}
\title{Level rearrangement in exotic-atom-like  three-body  systems}
\author{Jean-Marc Richard}
\email{j-m.richard@ipnl.in2p3.fr}
\author{Claude Fayard}
\email{c.fayard@ipnl.in2p3.fr}
\affiliation{Institut de Physique Nucl\'eaire de Lyon, Universit\'e de
Lyon, CNRS-IN2P3-UCBL, 4, rue Enrico Fermi, Villeurbanne, France}
\date{\today}
\begin{abstract}
We study systems of three bosons bound by a long-range interaction supplemented by a short-range potential of variable strength. This generalizes the usual two-body exotic atoms where the Coulomb interaction is modified by  nuclear forces at short distances. The energy shift due to the short-range part of the interaction combines two-body terms similar to the ones entering the Trueman-Deser formula, and three-body contributions. A sudden variation of the energy levels is observed near the coupling thresholds of the short-range potential. But the patterns of rearrangement are significantly modified as compared to the two-body case.
\end{abstract}
\maketitle
\section{Introduction}
\label{sec:intro}
Exotic atoms have a long history, and have stimulated interesting developments in the quantum dynamics of systems involving both long-range and short-range forces. For the exploratory studies presented here, the refinements of effective theories~\cite{1999PhRvD..60k4030H} are not required, and we shall restrict ourselves to the Schr\"odinger framework, as reviewed, e.g., in~\cite{1996NuPhA.606..283G} for the three-dimensional case, and extended in \cite{2011JPhA...44A5302C} for the two-dimensional one.

In units simplifying the treatment of the pure Coulomb case,  where the  strength becomes 1 and the reduced mass $1/2$, an exotic atom can be modeled as
\begin{equation}\label{eq:intro1}
 -\Delta \Psi+ V\,\Psi=E\,\Psi~,\quad
 V=-\frac{1}{r}+ \lambda\,v(r)~,
\end{equation}
where $r$ is the inter-particle distance and $\lambda\,v(r)$ the short-range correction, with a variable strength for the ease of discussion. 

In most actual exotic atoms, there is a strong absorptive component in the short range interaction, so one has to use either a complex (optical) potential $v$ or a coupled-channel formalism. Probably, the $(\bar D_s,\,p)$ atom, with a proton and an anticharmed meson $\bar D_s=(\bar c s)$ of charge $-1$ and strangeness $-1$,  escapes any absorption, since it lies below any threshold such as $(\bar D^0(\bar c u)\,\Lambda(sud))$, but it is not yet accessible experimentally. In this  study, we make the somewhat drastic simplification of a purely real short-range term~$v$.

When the above spectral problem is solved, the most striking observations are:
\begin{enumerate}[leftmargin=.4cm]
\item The energy shift $\delta E=E-E_n$, as compared to the pure-Coulomb energies $E_n=-1/(4\,n^2)$, is often rather small, but is usually not given by ordinary perturbation theory: for instance, an infinite hard-core of small radius corresponds to a small energy shift but  to an infinite first order correction. 
\item  Each energy $E(\lambda)$, as a function of the strength parameter $\lambda$, is almost flat in a wide interval of $\lambda$, with a value close to some $E_n$, one of the pure Coulomb energies. For S-wave states $\delta E$ is well approached by a formula by Deser et al., and Trueman~\cite{1954PhRv...96..774D,*1961NucPh..26...57T}, 
\begin{equation}\label{eq:intro2}
 \delta E\simeq \frac{a}{2\, n^3}~,
\end{equation}
where $a$ is the scattering length in the short-range interaction $\lambda\,v(r)$ alone.

\item If $v(r)$ is attractive, when the strength $\lambda$ approaches one of the positive critical values at which a first or a new bound state appears in the spectrum of $\lambda\,v$ alone, the energy $E(\lambda)$ quits its plateau and drops dramatically from the region of atomic energies to the one of deep nuclear binding. It is rapidly replaced in the plateau by the next level. This is known as \emph{level rearrangement}~\cite{Zel59}. Note that level rearrangement disappears if absorption becomes too strong \cite{1996NuPhA.606..283G,2007IJMPB..21.3765C}.

\item  The above patterns are more general,  and hold for  any combination of a long-range and a short-range interaction, say $V=V_0+\lambda\,v(r)$, as encountered, e.g., in the physics of cold atoms where a long-range confining interaction is supplemented by the direct interaction among the atoms \cite{Busch1998}. The Deser-Trueman formula of S-states is generalized as 
\begin{equation}\label{eq:intro3}
 \delta E\simeq 4\pi\,|\Psi_0(0)|^2\, a~,
\end{equation}
where $\Psi_0$ is the normalized wave-function in the external potential~$V_0$.   
If $V_0$ supports only one bound state, then the rearrangement ``extracts'' states from the continuum, instead of shifting radial excitations that are already bound.  An illustration is given in Fig.~\ref{fig:rearr2bs}, with a superposition of two exponential potentials of range parameters $\mu=1$ and $\mu=100$, namely 
\begin{equation}\label{eq:pot-2b}
\begin{gathered}
 V(r)=2\, v_e(1,r)+\lambda\,v_e(100,r)~,\\
 v_e(\mu,r)=-1.4458\,\mu^2\,\exp(-\mu\,r)~,
 \end{gathered}
\end{equation}
where $v_e(\mu,r)$ is tuned to start binding at unit strength.
\begin{figure}[ht!]
\begin{center}
 \includegraphics[width=.29\textwidth]{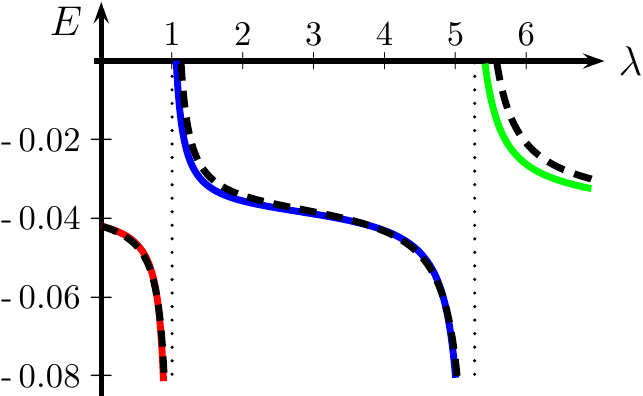}
\end{center}
 \caption{(Color online) Level rearrangement for a  two-body system bound by the potential~\eqref{eq:pot-2b}. The dashed line indicates the approximation corresponding to the Deser-Trueman formula~\eqref{eq:intro3}. The dotted vertical lines show the coupling thresholds for the short-range part of the interaction.}
 \label{fig:rearr2bs}
\end{figure}

\item  There are several possible improvements and alternative formulations of \eqref{eq:intro3}. For instance, $a$,  the bare scattering length in $\lambda\,v(r)$ can be replaced by the ``long-range corrected'' scattering length where the solutions of the radial equation for $\lambda\,v(r)$ are matched to the eigenfunction of the external potential. See, e.g., \cite{1983JPhG....9.1009T}, and refs.\ there. In the physics of cold atoms, one is more familiar with the approach by Busch et al.\ \cite{Busch1998}. It deals with the case of an harmonic oscillator modified at short distances, but the derivation can be  generalized  as follows. Let $u(E,r)=\sqrt{4\,\pi}\,r\,\Psi$ the $s$-wave reduced radial wavefunction for $V_0(r)$ that is regular at large $r$, at energy $E<0$, with some normalization, e.g., $u(E,r)\,\exp(r\sqrt{-E})\to 1$ for $r\to\infty$. The levels in $V_0$ correspond to the quantization condition $u(E_0,0)=0$, where $E_0$ is an eigenenergy of $V_0$, for instance the ground state.    When  a 
point-like interaction of scattering length $a$ is added, then the boundary condition is modified into 
\begin{equation}\label{eq:Busch1}
 u(E,0)+ a \,\partial_r u(E,0)=0~,
\end{equation}
which can be expanded near $E_0$ to give
\begin{equation}\label{eq:Busch2}
 \delta E=E-E_0\simeq - a \,\frac{\partial_r u(E_0,0)}{\partial_E u(E_0,0)}~.
\end{equation}
Now, the equivalence of \eqref{eq:Busch2} and \eqref{eq:intro3}, rewritten as \mbox{$\delta E\simeq a\, [\partial_r u_N(E_0,0)]^2$}, where $u_N$ is the normalized version of $u$,  comes from the relation
\begin{equation}\label{eq:Busch3}
\partial_r u(E_0,0)\,\partial_E u(E_0,0)=-\int_0^\infty u(E_0,r)^2\,\mathrm{d} r~,
\end{equation}
which is easily derived from the Wronskian identity, widely used in some textbooks \cite{Messiah:1140523}, here applied to energies $E$ and $E_0$.

\item The generalization to a number of dimension $d\neq 3$  is straightforward for $d>3$. For $d=1$, the first plateau is avoided, as the short-range potential, if attractive, develops its own discrete spectrum for any $\lambda>0$. The case of $d=2$ is more delicate: see, e.g., \cite{2007IJMPB..21.3765C,Zinner:2011ke}.
\end{enumerate}

Our aim here is to present a first investigation  of the three-body analog of exotic atoms. There are already studies of systems such as $(K^-,d)$, where $d=(p,n)$ is the deuteron, in which the neutron feels only the short-range part of the interaction. We will study systems in which the three constituents are already bound by the long-range component of the potential. this is the first attempt, at least to our knowledge. We consider three identical bosons, relevant for three atoms in a confining trap\footnote{At least for harmonic confinement, an external potential can be rewritten as a sum of pair interactions}. We  have in mind less symmetric systems for future work.  We address the following questions: Is there a pattern similar to the level rearrangement? Is there a generalization of the Deser-Trueman formula? What are the similarities with the case where the long-range interaction is replaced by an overall harmonic confinement?

Note that the occurrence of plateaus and sudden drops of the energies as a function of the coupling strength is not very usual, as these energies are monotonic and concave functions of any parameter entering linearly the Hamiltonian \cite{Thirring:2023497}. For instance, in a pioneering study of three-boson energies, Osborn \cite{osborn1967} fund some type of rearrangements in the three-body spectrum corresponding to a Yukawa interaction, 
but it was later acknowledged that this calculation suffers from some numerical instability, as the computed Faddeev energies violate a rigorous lower bound \cite{osborn1967,1968PhLB...27..195H}. Unfortunately, the erroneous plot was  reproduced in a seminal textbook on the three-body problem \cite{Schmid:239875}, 

The paper is organized as follows. In Sec.~\ref{sec:three-body}, we give some basic reminders about the spectrum of a three-boson systems from the Borromean limit of a single bound state to the  regime of stronger binding, with a word about the numerical techniques. 
 The results corresponding to a superposition $V_0(r)+\lambda \,v(r)$ are displayed in Sec.~\ref{sec:results}. An interpretation is attempted in Sec.~\ref{sec:interpretation}, with  a three-body version of the Deser-Trueman formula. Section~\ref{sec:outlook} is devoted to our conclusions.

\section{The three-boson spectrum  with a simple potential}\label{sec:three-body}
If two bosons interact through an attractive potential, or a potential with attractive parts, $\lambda\,v_0(r)$, a minimal strength is required to achieve binding, say $\lambda>\lambda_2^\text{cr}$. A collection of values of $\lambda_2^\text{cr}$ can be found, e.g., in the classic paper by Blatt and Jackson \cite{1949PhRv...76...18B}. In the following, we shall normalize $v_0$ so that $\lambda_2^\text{cr}=1$. 

If one assumes that $v_0(r)$ is attractive everywhere,  once two bosons are bound,  the $3$-boson system is also bound. (The case of potentials with a strong inner repulsion would require a more detailed analysis which is beyond the scope of this preliminary investigation.)\@
This means that for a single monotonic potential
\begin{equation}\label{eq:3b1}
 V_0=\lambda\left[v_0(r_{12})+v_0(r_{23})+v_0(r_{31})\right]~,
\end{equation}
where $r_{ij}=|\vec r_j-\vec r_i|$, the minimal coupling to achieve three-body binding, $\lambda=\lambda_3^\text{cr}$, is less than 1, in our units. This is implicit in the seminal paper by Thomas \cite{1935PhRv...47..903T}. The inequality $\lambda_3^\text{cr}<\lambda_2^\text{cr}$ is now referred to as ``Borromean binding'', after the study of neutron halos in nuclear physics~\cite{1993PhR...231..151Z}. One gets typically $\lambda_3^\text{cr}\simeq 0.8$, i.e., about 20\% of Borromean window~\cite{1994PhRvL..73.1464R}. 

\begin{figure}[!htb]
\begin{center}
 \includegraphics[width=.3\textwidth]{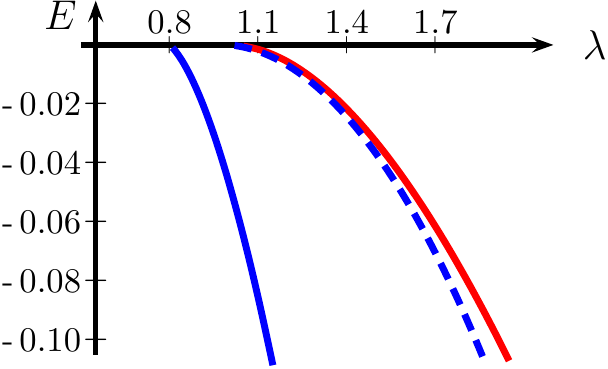}
 \\[10pt]
 \includegraphics[width=.3\textwidth]{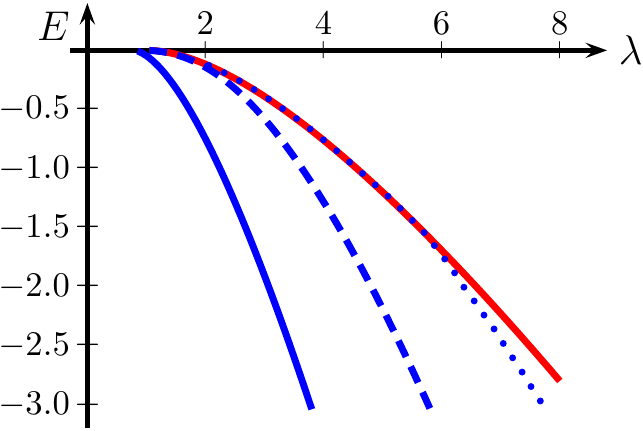}
\end{center}
 \caption{(Color online) Two-boson (red) and three-boson spectrum (blue) in the potential \eqref{eq:3b1}, with $v_0(r)\propto -\exp(-r)$ at small coupling (top) and larger coupling (bottom). The first excited 3-body state (dashed) is always found below the 2-body energy. The second-excited state (dotted line in the bottom plot) and higher states (not shown) require a minimal coupling to be bound below the dissociation threshold. \label{fig:levels3b} } 
\end{figure}

In short, the three-boson spectrum has the following patterns:
\begin{itemize}
 \item For $\lambda<\lambda_3^\text{cr}$, no binding
 \item For $\lambda_3^\text{cr}<\lambda<1$, a single Borromean bound state, and, for $\lambda$ close to 1,  a second three-body bound state just below the two-body energy, 
 \item Very near $\lambda=1$, the very weakly bound Efimov states. 
 \item For $1<\lambda$, two bound states below the $2+1$ break-up, and further bound states when $\lambda$ becomes very large. 
\end{itemize}
These patterns are independent of the detailed shape (exponential, Gaussian, \dots) of the potential $v_0(r)$ , if monotonic. For potentials with an internal or external repulsive barrier, one expects some changes, as already the width of the Borromean window is modified \cite{PhysRevA.62.032504}.
This is  illustrated in Fig.~\ref{fig:levels3b} for the case of an exponential potential. 

The first excited state can be considered as the first member of the sequence of the Efimov states occurring near $\lambda=1$. However, it differs from the other Efimov states in the sense that when the coupling is increased, it remains below the two-body break-up threshold, at least for the simple monotonic potentials we consider here. \footnote{We thank Pascal Naidon (RIKEN) for a correspondence on this point.}

The two-body energy is know analytically for a single exponential potential. For the three-body energies, we first made some investigations based on the hyperspherical expansion \cite{1998FBS....25..199K}. The results were finalized  with a variational method based on either exponential of Gaussian wavefunctions, say%
\begin{equation}\label{eq:psi}
 \Psi=\sum_i \gamma_i \left[\exp(-a_i\, r_{23}^n-b_i\,r_{31}^n-c_i\,r_{12}^n)+ \cdots\right]~,
\end{equation}
where the dots stand for terms deduced by permuting the particles. For a given choice of range parameters $a_i$, $b_i$ and $c_i$, the Schrödinger equation results in a generalized eigenvalue equation, whose eigenvectors are the coefficients $\{\gamma_1,\gamma_2,\ldots\}$ and eigenvalues upper (variational) bounds on the exacts energies. The range parameters are thus tuned to minimize the energy levels.
For $n=1$, we wrote our own code: the range parameters $a_i$, $b_i$ and $c_i$, if restricted to be real, are chosen to belong to  a geometric progression $\{\alpha,\, \alpha\,r, \alpha\,r^2, \ldots\}$ of common ration $r>1$ and scale factor $\alpha$. The lowest term $\alpha$ can be linked to the energy $E$ by the relation $\alpha^2=14\,E/15$ suggested by the Feshbach-Rubinow equation~\cite{1955PhRv...98..188F}. For excited states, we found it more efficient to introduce complex range parameters $a_i$, $b_i$ and $c_i$ as done, e.g., by Korobov \cite{2000PhRvA..61f4503K} for three charge ions in atomic physics, and to adjust the  $a_i$, $b_i$ and $c_i$  by random trials. For $n=2$, we used the code made available by Suzuki and Varga~\cite{1997CoPhC.106..157V}, with minor changes. 
Anyhow, our aim was not to produce very accurate benchmark energies, but to identify the main patterns. 

\section{The three-body energies with long- and short-range forces} \label{sec:results}
We now replace \eqref{eq:3b1} by a superposition
\begin{equation}
 \label{eq:3b2}\sum_{i<j}\left[ \lambda_0\,v_e(1,r_{ij})+ \lambda\,v_e(\mu,r_{ij})\right]~,
\end{equation}
where $v_e(\mu,r)$ is significantly shorter ranged than the external potential $v_e(1,r)$. In practice, we will choose  $\mu$ ranging from 10 to 30.  Note that the computations become rather delicate for larger values of $\mu$, and would require dedicated techniques. 

\begin{figure}[hb!]
\begin{center}
\includegraphics[width=.25\textwidth]{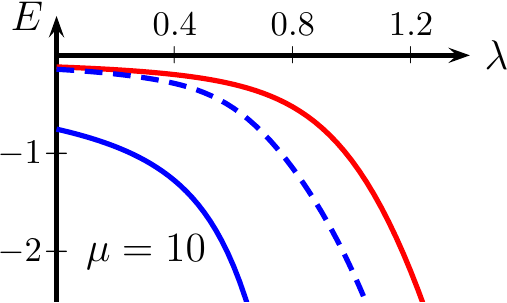}
 \\[8pt]
\includegraphics[width=.25\textwidth]{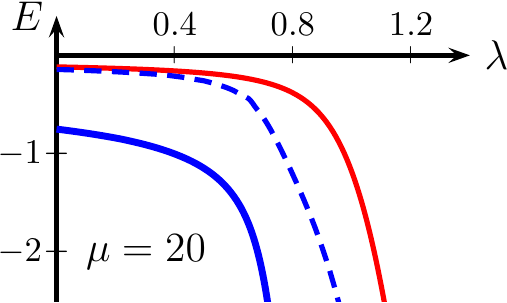}
 \\[8pt]   
\includegraphics[width=.25\textwidth]{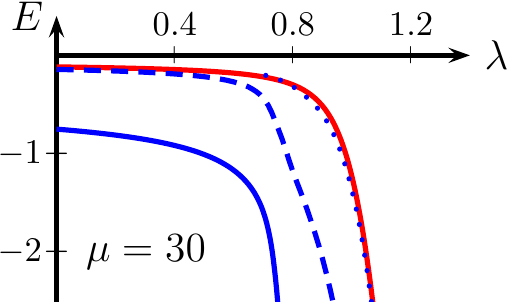}
\end{center}   
 \caption{(Color online) Spectrum in the potential \eqref{eq:3b2} for (top to bottom) a ratio of ranges $\mu=10,\,20$ and 30, as a function of the strength of the short-range potential. Red: Two-body energy, Blue: three-body levels, solid line for the ground state, dashed for the first excitation, dotted for the second excitation, visible only for $\mu=30$  (for smaller $\mu$, the second excited state requires a coupling $\lambda$ outside of the range of the plot)}\label{fig:2e}
\end{figure}

In Fig.~\ref{fig:2e} are shown the spectra for  $\lambda_0=2$, i.e., twice the two-body critical coupling and varying $\lambda$, for  $\mu=10,\,20$ and $30$.
The unit of energy is irrelevant, as it can be modified by an overall rescaling of the distances. Comments are in order:
\begin{itemize}
\item As in Fig.~\ref{fig:rearr2bs}, a convex behavior  as a function of $\lambda$ is observed for the  excited energy-levels, i.e.,  $E_n(\lambda)$ with $n>1$. This is permitted, provided that the sum of the first energies remains concave~\cite{Thirring:2023497}. 
 \item As in the two-body case, the transition is sharper when the range of the additional potential becomes shorter.
 \item There is clearly a beginning of rearrangement, in the sense that for $\lambda\to \lambda_3^\text{cr}\simeq 0.8$, the excited state falls suddenly near the unperturbed ground-state energy.
 \item However, there is no second plateau for the excited state, just somewhat a smoothing of the fall-off for~$\lambda \gtrsim \lambda_3^\text{cr}$, together with a tendency to smooth or to invert the concavity. 
\item 
The second excited state, of energy $E_3^{**}$, becomes bound (i.e., below the energy $E_2$ of its dissociation threshold) for smaller and smaller values of the coupling $\lambda$. We get  typically a coupling threshold of about 5 for a single exponential, about 1 for $\mu=10$,  and  about $\lambda=0.5$ for $\mu=30$. The binding energy $E_2-E_3^{**}$ has seemingly some delicate pattern, as it is not monotonic as a function of $\lambda$, at least in our calculations. We cannot exclude that for another choice  of potentials, the  second excited state becomes bound, then is reabsorbed by the two-body threshold, and eventually reappears for larger $\lambda$. In this scenario, the number of normalizable three-body bound states below the two-body threshold would not be a monotonic function of the strength $\lambda$, as in the Efimov effect. 
\end{itemize}
\section{Interpretation}\label{sec:interpretation}
Let us first concentrate on the region of small $\lambda$. One can estimate the energy shifts $(\delta E)_{ij}$ corresponding to  several external potentials $V_{0,i}$ with $i=1, \ldots N$ and short-range potentials $V_j$,  with $j=1, \ldots N'$ and $N,N'>3$, and study empirically the properties of the matrix $\{(\delta E)_{ij}\}$. 

In the two-body case, one finds that the $2\times 2$ sub-determinants vanish almost exactly. This is compatible with a factorization 
\begin{equation}\label{eq:3b3}
 \delta E^{(2)}\simeq A_\text{LR}\,B_\text{SR}~,
\end{equation}
as a product of a long-range term depending only on $v_0$ and a short-range term depending only on $\lambda\,v$. This factorization is achieved by the Deser-Trueman formula, with $A_\text{LR}$ being the square of the wave function at $r=0$ (times $4\,\pi$) and $B_\text{SR}$ the scattering length. 

\begin{figure}[ht!]
 \includegraphics[width=.3\textwidth]{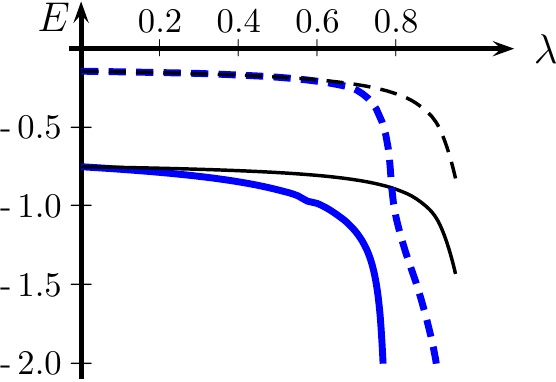}
 \caption{ (Color online) Calculated energy (thick blue line, dashed for the excited state) vs.\ the estimate using the first term in Eq.~\eqref{eq:3b4} (thin black line). The calculation is done for a superposition of two exponentials, one with unit range and strength $\lambda_0=2$, and another of range parameter $\mu=50$ and strength $\lambda$.}
 \label{fig:trueman3}
\end{figure}

In the three-body case, it is observed that the $2\times 2$ sub-determinants still nearly vanish, especially for the smaller values of the short-range strength $\lambda$, but that the  $3\times 3$ sub-determinants vanish even better (of course we compared the determinants divided by the typical values of a product of 2 or 3 $\delta E$). This is compatible with $\delta E$ being a sum of two factorized contributions,
\begin{equation}\label{eq:3b4}
 \delta E^{(3)}=A_\text{LR}\,B_\text{SR}+A'_\text{LR}\,B'_\text{SR}~.
\end{equation}
As explained, e.g., in the textbook by Ericson and Weise \cite{Ericson:112241}, the Deser-Trueman formula \eqref{eq:intro3} can be understood as the perturbative correction due to a Fermi point-potential that includes non perturbatively the effect of the short-range  interaction. Thus for  a symmetric three-body system, the same prescription leads to
 a simple extension of \eqref{eq:intro3} that  reads
\begin{equation}\label{eq:3b5} 
A_\text{LR}\,B_\text{SR}=12\,\pi\,|\Psi_{12}(0)|^2\,a~,
\end{equation}
where $|\Psi_{12}(0)|^2$ is a short-hand notation for the two-body correlation factor $\langle \Psi |\delta^{(3)}(\vec r_2-\vec r_1)|\Psi\rangle/\langle\Psi|\Psi\rangle$. It is checked that this term dominates for small shifts, i.e., for small $\lambda$, see Fig.~\ref{fig:trueman3}. However, this term  alone would induce a sharp decrease of the atomic energies only for $|a|\to\infty$, i.e., for $\lambda\to 1$, the coupling threshold for two-binding, and not near $\lambda=0.8$, as actually observed.  For an asymmetric three-body system, the extension is, in an obvious notation,
\begin{equation}\label{eq:3b5a}
3\,\pi\,\sum_{i<j}|\Psi_{ij}(0)|^2\,a_{ij}~.
\end{equation}

The second contribution in Eq.~\eqref{eq:3b5} should thus account for the genuine three-body effects, besides some higher-order two-body terms such as a contribution from effective range.  The three-body part of $B'_\text{SR}$ is a kind of generalized scattering length that blows up when $\lambda$ approaches the coupling threshold $\lambda_3^\text{cr}$ for three-body binding. As the theory of three-body scattering is a little intricate, we postpone the precise definition of $B'_\text{SR}$ to some further study. As for the long-range factor $A'_\text{LR}$ of this second term, the simplest guess is to assume that it is proportional to the square of the wavefunction at $\vec r_1=\vec r_2=\vec r_3$, or in terms of the Jacobi variables $\vec x$ and $\vec y$ describing the relative motion, $A'_\text{LR}\propto 
\langle \Psi |\delta^{(3)}(\vec x)\,\delta^{(3)}(\vec y)|\Psi\rangle$, but this is seemingly not the case.

Our study of generalized exotic atoms is related to the Efimov physics. 
In particular, the authors of refs.~\cite{2008PhRvA..78b0501T,2011FBS....51..219P}, and probably  some others, have studied how the Efimov effect is modified if each atom is submitted to an individual harmonic confinement.  They also found that near a point where the two-body scattering length becomes infinite, there is a finite number of three-body bound states, instead of an infinite number in absence of confinement.  The second-excited three-body bound state in  Fig.~\ref{fig:2e} is slightly reminiscent of  an Efimov state of the short-range potential, modified by the long-range potential. 

\section{Conclusions}\label{sec:outlook}
The lowest states of three-bosons have been calculated with a superposition of  long-range and  short-range attractive potentials. When the strength $\lambda$ of the latter is increased, starting from $\lambda=0$, the three-body energies decreases very slowly, and can be well approximated by a straightforward generalization of the Deser-Trueman formula involving only the two-body scattering length. However, when $\lambda$ approaches 0.8 (in units where $\lambda=1$ is the coupling threshold for binding in the short-range potential alone), there is a departure for the Deser-Trueman formula, which can be empirically accounted for by the product of a short-range and a long-range factor. The short-range factor is the three-body analog of the scattering length and becomes very large when $\lambda\simeq 0.8$ which corresponds to the occurrence of a Borromean bound state in the short-range potential alone. 

This exploratory investigation has been done with a simple variational method based on a few exponential or Gaussians functions, which  is sufficient to show the main trends. More powerful minimization methods are probably required for the states at the edge of stability with respect to spontaneous dissociation. We also studied some modeling with separable potentials. The qualitative patterns of rearrangement are observed for separable potentials of rank 2 or higher, with some slight differences with respect to the case of local potentials. This will be presented in a forthcoming article. 

Many other developments are required. What is the precise definition of the three-body short-range factor? What is the corresponding long-range factor? What is the minimal ratio of range parameters required for the occurrence of the third stable three-body state? When does a fourth state show up? What are the analogs for $N\ge 4$ bosons? We also aim at studying some asymmetric systems. For instance, a prototype of $(K^-pp)$ could be built, with a Coulomb interaction, that is known to produce a stable ion, below the threshold for breakup into a $(K^-p)$ atom and an isolated proton~\cite{1977JMP....18.2316H}. Then the strong interaction between the two protons and the strange meson $K^-$ could be mimicked by a simple potential of range about $1\,$fm,  first real, and then, more realistically, complex to  include absorptive effects, to study how the existence of a nuclear bound state $(K^-pp)$  modifies the atomic spectrum.
\begin{acknowledgments}
 One of us (JMR) would like to thank ECT*, Trento, for the hospitality extended to him.
\end{acknowledgments}
\clearpage
%
%

\end{document}